\documentclass[journal]{IEEEtran}
\ifCLASSINFOpdf
\else
\fi
\usepackage{graphicx}
\usepackage{cite}
\usepackage{algorithm}
\usepackage[noend]{algpseudocode}
\usepackage{algorithmicx}
\usepackage{hyperref}

\usepackage{amsmath}
\usepackage{amssymb}
\begin{document}
%
\title{Towards Intent-Based Network Management: Large Language Models for Intent Extraction in 5G Core Networks
}
%
%
%

\author{Dimitrios~Michael~Manias, Ali~Chouman, and~Abdallah~Shami\\ The Department of Electrical and Computer Engineering, Western University\\ \{dmanias3, achouman, Abdallah.Shami\}@uwo.ca 
\thanks{© 2024 IEEE.  Personal use of this material is permitted.  Permission from IEEE must be obtained for all other uses, in any current or future media, including reprinting/republishing this material for advertising or promotional purposes, creating new collective works, for resale or redistribution to servers or lists, or reuse of any copyrighted component of this work in other works.}
}

%
%

\markboth{International Conference on the Design of Reliable Communication Networks}%
{Manias \MakeLowercase{\textit{et al.}}:Towards Intent-Based Network Management: Large Language Models for Intent Extraction in 5G Core Networks}
%



\maketitle

\begin{abstract}
The integration of Machine Learning and Artificial Intelligence (ML/AI) into fifth-generation (5G) networks has made evident the limitations of network intelligence with ever-increasing, strenuous requirements for current and next-generation devices. This transition to ubiquitous intelligence demands high connectivity, synchronicity, and end-to-end communication between users and network operators, and will pave the way towards full network automation without human intervention. Intent-based networking is a key factor in the reduction of human actions, roles, and responsibilities while shifting towards novel extraction and interpretation of automated network management. This paper presents the development of a custom Large Language Model (LLM) for 5G and next-generation intent-based networking and provides insights into future LLM developments and integrations to realize end-to-end intent-based networking for fully automated network intelligence.
\end{abstract}


%
\IEEEpeerreviewmaketitle

\section{Introduction}
%
%
%
%
\IEEEPARstart{T}{he} fifth generation of networking, commonly referred to as 5G, has revolutionized network operations worldwide. Despite still being in development, initial deployments have begun to emerge globally. The advent of this network generation has put increasing pressure on network service providers due to the increasingly stringent performance requirements needed to ensure that the true potential of these networks can be harnessed. For instance, 5G networks offer increased user connection density, increased speeds, and reduced latency \cite{navarro2020survey}. Research into the next generation of networking, 6G, has already begun. To this end, understanding the limitations of current 5G networks and the steps required to transition into 6G networks is critical. \par

5G networks are the first networking generation to integrate Machine Learning and Artificial Intelligence (ML/AI). This integration is found in the core network, namely the 5G Core Network Data Analytics Function (NWDAF) \cite{chouman2022towards}. This function aims to act as a centralized intelligence agent capable of monitoring, analyzing, predicting and prescribing information based on network-generated data. In the realm of networking, the development of this function is an essential first step into the integration of intelligence into the network; however, in order to address the limitations of 5G networks and move towards the Beyond 5G (B5G) era of networking, a much more profound integration of intelligence must be woven into the fabric of the network itself. This suggests moving past a single network function responsible for intelligence into a paradigm-shifting model of ubiquitous intelligence. This ubiquitous intelligence would entail an end-to-end synchronized, connected, and autonomous system capable of interfacing with various entities in the network, such as the client, the user, the operator, the service provider, and the manager, to name a few. It should be noted that the proliferation of intelligence in networks gives rise to additional operational challenges that must be considered as part of its integration, such as the manifestation of model drift in highly dynamic network settings \cite{10163748}. \par

This paradigm shift allows the realization of full network automation to the prescribed level in Zero-Touch Network Service Management (ZSM) architectures. The ZSM architecture defines a network with qualities such as self-healing, self-configuration and self-optimization \cite{ZSM}. In order to attain these qualities, networks must constantly sense and interpret the required changes to maximize performance. With the introduction of network slicing, essentially defining logical isolation in virtualized networks, a large part of the network management relates to the end user and the application being requested. As such, a fully autonomous network system should be able to interpret the user's intentions and configure the network accordingly. \par

Furthermore, all network configurations and reconfigurations should be communicated to an intelligent agent to reduce human intervention and the chance of a human-induced error. This agent should be able to interpret what is required and act accordingly. In the envisioned system, network operation will be based on extracting a set of intents and converting them into actions and policies autonomously and without human intervention. Instead, the human's role will be strictly limited to supervision and oversight. \par

The notion of intent interpretation for automation introduces a second paradigm shift known as intent-based networking. As previously discussed, in this paradigm, the intent of a user or operator is extracted and converted into an actionable policy that can be enacted directly through a network controller. The first step towards achieving this intent-based networking revolves around the ability to extract intent from user input. This input can be voice, text, or any other form of information and must be interpreted accurately and consistently by the network. To this end, the work outlined in this paper takes a step in this direction by discussing the use of Large Language Models (LLMs) for intent extraction and interpretation in B5G core networks.\par

Recently, LLMs have taken the ML/AI space by storm. Their introduction has propelled the field of generative AI into a new and previously undiscovered frontier. These models are positioned to be prime candidates for intent extraction and conversion in next-generation networks due to their ability to efficiently and consistently generate high-quality pieces of text, code, and other forms of writing, as well as interpret the user's intentions. To this end, the work presented in this paper leverages these LLMs for core-based intent extraction related to network management operations. \par

The contributions of this paper can be summarized as follows:

\begin{itemize}
\item The development of a customized Large Language Model for intent extraction in 5G and Beyond core network operations
\item An insight into the state of Large Language Models and their future developments
\item A discussion of the steps required to extend the work in this paper into an end-to-end autonomous intent-based networking architecture
\end{itemize}

\section{Related Work}

Intent-based networking has gained significant traction in recent years. Leivadeas and Falkner \cite{leivadeas2022survey} present a comprehensive survey outlining the state-of-the-art in the field. The authors discuss the closed-loop automation aspect through intents with stages such as intent profiling, translation, resolution, activation, and assurance. Their in-depth analysis outlines methods such as Natural Language Processing (NLP), which can transform an intent into an actionable policy. The authors also present a set of open challenges in the field, including Zero Touch Networks and industry-specific intents. Velasco \textit{et al.} \cite{velasco2021end} discuss the topic of end-to-end intent-based networking. The authors present an ML-based framework that monitors the network condition and enacts management and orchestration decisions through a programmable data plane according to ML-generated intents. Njah \textit{et al.} \cite{njah2023toward} present an NLP-based approach to intent-based networking, specifically showcasing a healthcare use case. The objective of the work is to convert an unstructured intent to a structured intent, which includes properties such as the user, the application goal, the network action, the target equipment, and the time frame. The authors compare their solution to the state-of-the-art solutions and demonstrate a significant performance improvement when using a standard industry-specific benchmarking dataset. Wang \textit{et al.} \cite{wang2023network} discuss the integration of ChatGPT with autonomous network management and control. The authors develop an LLM-based model that is used to extract insights from network packet data. The authors identify use cases such as customer service, distributed task automation, and fraud detection as possible applications for their work. 

Intent-based networking is not limited to core network management, but can also aid in network slice management, configuration and network policies for decision-making \cite{da2023integration}. Abbas \textit{et al.} \cite{abbas2023ai, abbas2021network} discuss the use of intent-based networking for network slice automation. The authors assert the need for automation to move beyond manual configuration for network slicing. In their framework, users specify QoS requirements, which are then translated into network slice templates. Mcnamara \textit{et al.} \cite{mcnamara2023nlp} consider the use of NLP for intent-based network management specifically applied to private 5G Networks; NLP has garnered attention in recent works as it can provide a model framework for such private networks \cite{wang2021intent}. The authors present an intent engine based on adaptive policy execution, including intent handling, matching, and action building. The authors present various intent-related workflows and identify use cases such as slice management, line of sight channel identification, and service provisioning as potential applications for their work. Wei \textit{et al.} \cite{wei2020intent} present a set of insights and challenges into the use of intent-based networking in B5G networks. The authors outline the key technologies required for next-generation intent-based networking. They also discuss the future of intent-based networking and envision a more profound AI integration as future networks begin to take shape and materialize. \par

Most of the presented works focus on intent-based networking without harnessing the true potential of LLMs. These models can understand conversational text and extract relevant information to transform what the user is requesting into an actionable and interpretable policy for the network. One characteristic of this work that distinguishes it from the rest is that it focuses on a very specific task, intent extraction in the 5G core Network, something which, to the best of our knowledge, has yet to be considered thus far. This work aligns with the most recent standardization, and the intents discussed are derived from the most recent Third Generation Partnership Project (3GPP) standards. This work is the start of a bottom-up approach to end-to-end intent-based networking, leading to true network automation and ZSM. By adopting a microfunctionality approach and addressing a specific task in this work, we ensure system modularity and have the power to harness increased performance through selective and specific fine-tuning in the future. \par

\section{Background}
The 3GPP has released a technical specification related to the Management and Orchestration, specifically Intent-Driven Management Services for Mobile Networks \cite{intent_standard}. This technical specification defines a set of intents containing an expectation for the 5G core network. These intents form the foundation of this work and can be summarized as follows:
\begin{itemize}
    \item Deployment Intent
    \item Modification Intent
    \item Performance Assurance Intent
    \item Intent Report Request
    \item Intent Feasibility Check
    \item Regular Notification Request
\end{itemize} \par

Each of these defined intents serves a specific purpose in the grand scheme of intent-based networking for the 5G core network. Despite being based on the 5G core network, the work presented in this paper applies to any future networking generation with clearly defined intents. In terms of the 5G core network, the deployment intent is invoked every time a new 5G core network is required in a specified area. Parameters related to this employment intent can include a geographic location, the type of network, PLMN information, and target network capacity information, to name a few. The modification intent is used to modify an existing 5G core network based on expectations outlined in the intent. The performance assurance intent is used to prescribe a specific performance expectation for a 5G core network. Some parameters that can be used to define this intent include the required number of registered users or the number of created PDU sessions, to name a few. The intent report request is a unique type of intent that leads to information retrieval and details about the status of previously expressed intents. Information and details that can be requested about intents include achieved vs. target values, feasibility check information, conflict information, and fulfillment status. As its name suggests, the intent feasibility check assesses the feasibility of an expressed intent. This check is an intent and ensures that other intents have the available resources, capacities, and capabilities to be practically applied through the associated policies. The final intent is the regular notification request, which essentially subscribes a user to receive updates on the Fulfillment status of an intent.\par

It should be noted that a single request from a user, operator, or service provider can contain multiple intents. This means that it is the job of the agent first to recognize the existence of multiple intents, then interpret their semantic meaning, and finally invoke the appropriate procedures to execute those requests. Some basic examples of the structure of the six outlined intents are listed in Table \ref{intents}. \par

\begin{table}[]
\caption{5G Core Intent Types and Examples}
\label{intents}
\begin{tabular}{|p{3.5cm}|p{4.5cm}|}
\hline
\textbf{Intent Type}         & \textbf{Intent Structure Example}                                                                               \\ \hline
Deployment Intent            & “Deploy a new network in {[}region{]} with the following specifications...”                                     \\ \hline
Modification Intent          & “Modify the existing {[}network{]} to address the performance issues caused by high loading...”                 \\ \hline
Performance Assurance Intent & “Ensure that the deployed network can support a {[}QoS Level{]} application with the following requirements...” \\ \hline
Intent Report Request        & “Summarize the results of the previous request.”                                                                \\ \hline
Intent Feasibility Check     & “Before proceeding, ensure that capacity exists in {[}region{]} to perform the required changes.”               \\ \hline
Regular Notification Request & “Notify me of the status of {[}network{]} every {[}frequency{]}.”                                               \\ \hline
\end{tabular}
\end{table}

After defining the appropriate intents, the next stage is integrating them with an LLM that will transform the input request through intent extraction into an actionable policy that can be directly enacted in the network. For this preliminary work, the LLM of choice has been OpenAI’s ChatGPT 3.5. GPT 3.5 is one of the flagship state-of-the-art LLMs that exist today. It leverages transformer architecture and is pre-trained on massive volumes of internet-available text-based data. This model is an example of generative AI as it generates text responses based on a provided input. This model was selected mainly for its availability and versatility, as it can be customized for use in specific tasks through techniques such as prompting. One of the limitations of this model is that it is closed-source, meaning that the specific architecture and training process is not publicly available. Future iterations of this work will leverage state-of-the-art open-source LLMs, which provide greater flexibility and insight into the model, data, and training process.

\section{Methodology}

\begin{figure*}[!htbp]
\centerline{\includegraphics[width=1.5\columnwidth]{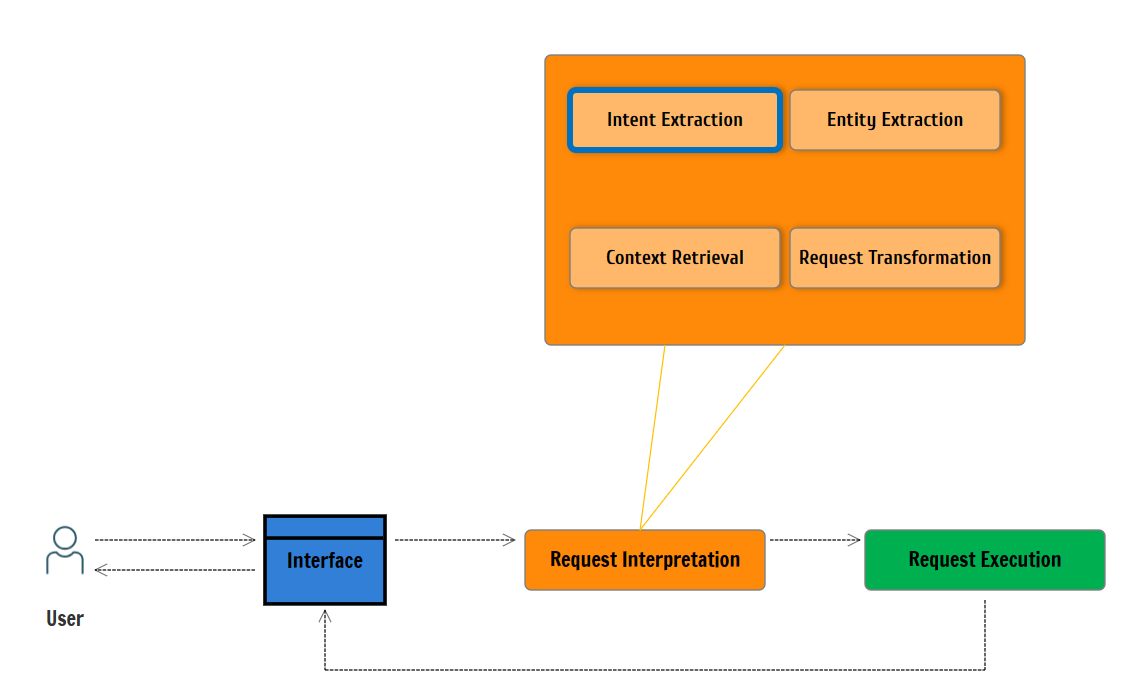}}
\caption{High-Level System Component Overview}
\label{overview}
\end{figure*}

Figure \ref{overview} depicts a high-level general architecture for intent-based networking. This architecture has four main components: the user, an interface, the request interpretation module, and the request execution module. When the user sends a request, it goes to the interface. This interface is responsible for two-way communication with the user as it receives user input and outputs responses back to the user. Once the interface receives the request, it gets routed to the request interpretation module. This module contains many components, some of which include intent extraction, entity extraction, context retrieval, and request transformation, to name a few. The work presented in this paper is specifically concerned with the intent extraction functionality of the request interpretation module. The purpose of the request interpretation module is to take a user request and get all the relevant information in order to format the request in a way that it can be enacted and executed in the network. This information extraction is done by identifying the components of the request, such as the entity and the intent, getting all the necessary context required in order to interpret what is being asked by the user and any supporting information that is required, and finally, transforming the request in a way that the network can interpret as a direct policy such as a JSON-formatted document. Once formatted, the request is passed along to the request execution module, which calls all appropriate APIs and ensures that the appropriate steps are taken to complete the request. This module reports back to the user when a request is completed. \par

For this work, prompting was the technique of choice for customizing the functionality and output of the LLM. Specifically, the prompt includes a role, a description of the task, background context from the standards providing a description of the intents, and a description of the expected behaviour. Using such an architecture in the prompting phase enables the model to understand precisely its task and how it should respond. Table \ref{prompting} outlines the various prompt components as implemented. \par

\begin{table}[]
\caption{Prompting Architecture Components}
\label{prompting}
\begin{tabular}{|p{3.5cm}|p{4.5cm}|}
\hline
\textbf{Prompt Component} & \textbf{Implementation}                                                                                                                                                                                                                                                                                                                         \\ \hline
Role                      & “You are an intelligent agent within the 5G Core network.”                                                                                                                                                                                                                                                                                      \\ \hline
Task Description          & “Your task is to classify user intents into 6 categories. The categories are: Deployment Intent, Modification Intent, Performance Assurance Intent, Intent Report Request, Intent Feasibility Check, Regular Notification Request.”                                                                                                             \\ \hline
Background Context        & Excerpt from technical standard: 3GPP TS 28.312 V18.1.1 (2023-09)                                                                                                                                                                                                                                                                               \\ \hline
Expected Behaviour        & “Each request can have multiple intents. Your job is to specify which intents are present in each user request. If there is no intent present or you do not understand, please return “no intent present” or “unknown intent”; otherwise return all the intents that are present with an explanation as to why you have selected those intents” \\ \hline
\end{tabular}
\end{table}

It should be noted that explainability has been considered in the expected behaviour section of the prompt as the model is asked to justify why it believes a specific intent is present. Explainability is a critical component of any AI deployment involving decision-making and critical services. Its trustworthiness is questionable without understanding why a model has converged to a result. To this end, giving it the responsibility to contribute to decision-making processes in large-scale systems, without proper explanation, is imprudent. This initial first step towards explainability clarifies why the specific model prompted with this specific task given specific context has led to the returned response. This level of explainability is an advantage of using large language models and can also be used to tune the performance down the line. If the model returns a result with an explanation that is not sound or consistent with what is expected, a fine-tuning dataset giving the correct response or the expected response can be constructed, and the model can be fine-tuned to eliminate the inconsistency. \par

\section{Results and Analysis}
In order to evaluate the quality of the developed model, representative examples of various intents and non-intents were passed to the model as a request. These examples cover many scenarios, including single intents, multiple intents, and no intents present in a single request. The results of some of these scenarios are shown below to illustrate the ability of the LLM through prompting to accurately determine which intents are present, if any, and provide an appropriate justification as to why.

\begin{figure}[!htbp]
\centerline{\includegraphics[width=\columnwidth]{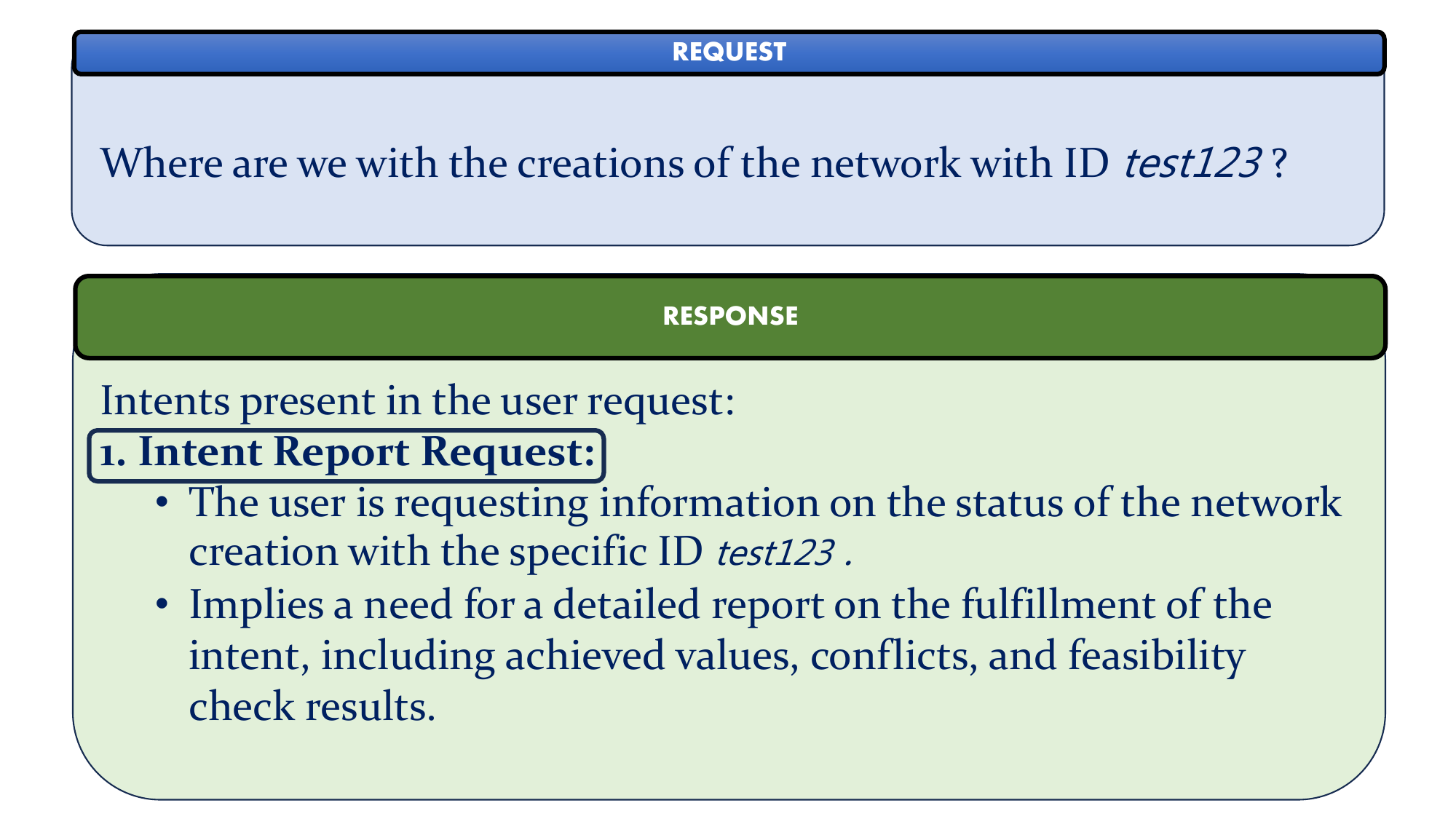}}
\caption{Single Intent Example 1}
\label{single-1}
\end{figure}

Figure \ref{single-1} presents a request with a single intent. In this request, the model is asked to provide a report on the status of a previous request. It should be noted that no context was provided as to the existence of a previous request. In the response, the LLM has accurately determined that the semantic meaning of the current request is in relation to a previous request and has accurately identified that the intent present is the intent report request. In its explanation, it accurately identifies that information on the status of a network creation intent is being requested. It also explicitly states that an implication related to a previous intent exists in the current request.

\begin{figure}[!htbp]
\centerline{\includegraphics[width=\columnwidth]{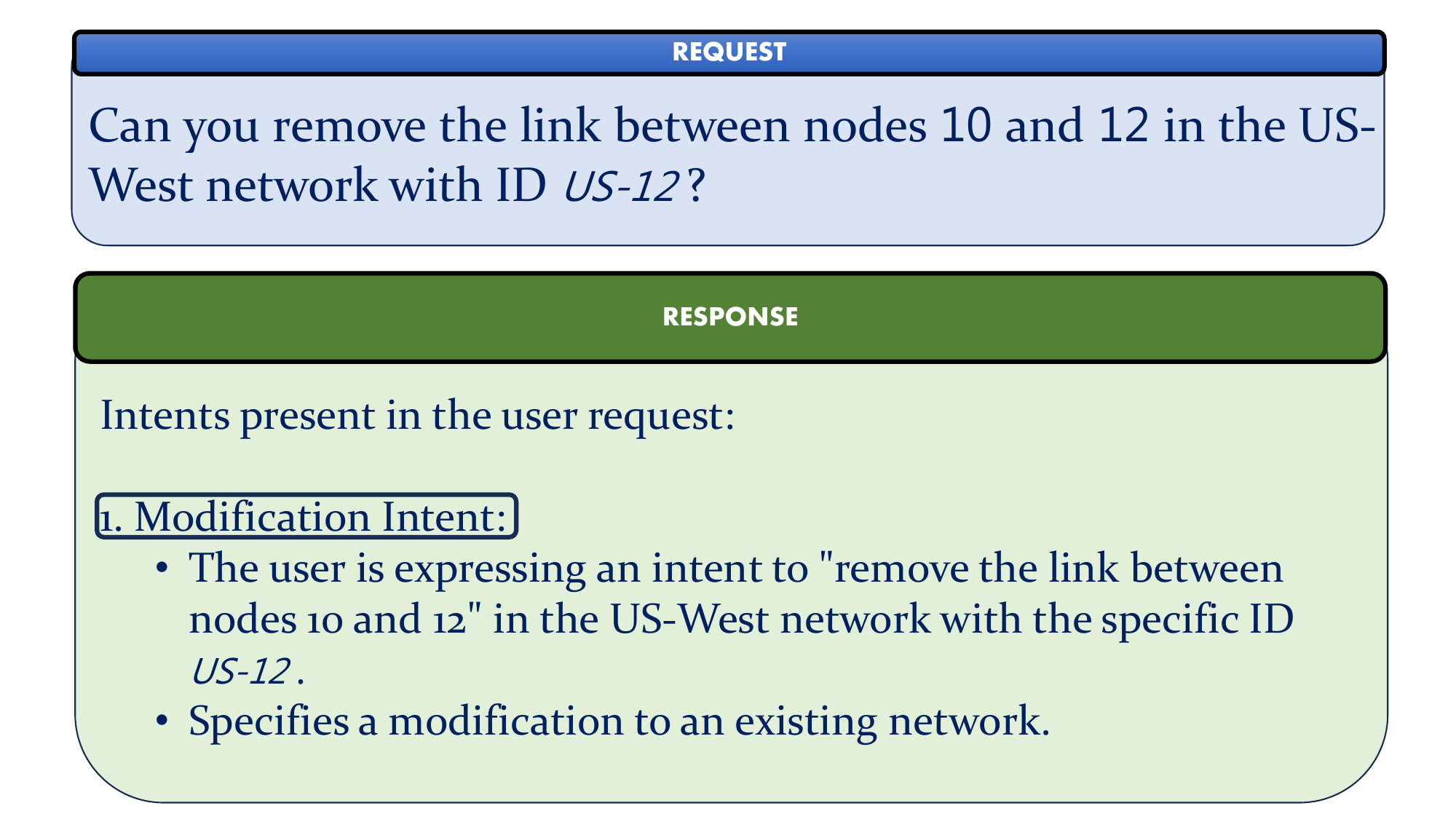}}
\caption{Single Intent Example 2}
\label{single-2}
\end{figure}

Figure \ref{single-2} presents another request with a single intent present. In this example, a modification to an existing network is requested. In the response, the modification intent has been correctly identified, and the supporting explanation demonstrates the ability of the LLM to identify the critical components of the request, such as what is being removed and from where. To humans, this might seem like a trivial request; however, for a machine to interpret and extract the multiple components and the semantic meaning of this request is no trivial task. This example highlights, at a fundamental level, the power of LLMs in today's world and their ability to transform networking practices and capabilities.

\begin{figure}[!htbp]
\centerline{\includegraphics[width=\columnwidth]{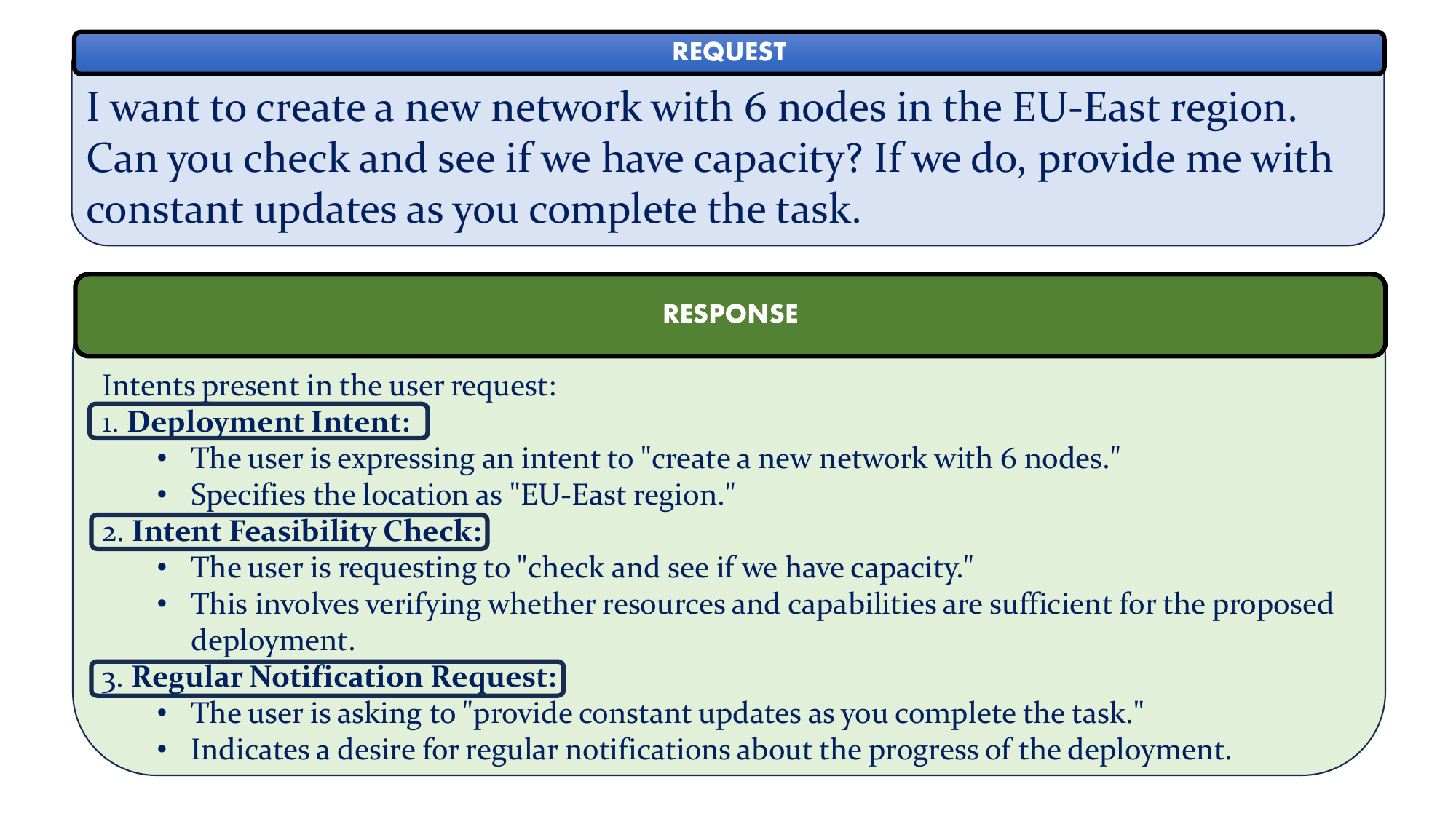}}
\caption{Multiple Intents Example 1}
\label{multi-1}
\end{figure}

Figure \ref{multi-1} presents an example of a more complicated request with multiple intents. In the prompting structure, it was explicitly communicated to the model that it should be identifying all the intents present. In conventional machine learning, this would require a multi-class, multi-label classification problem with a labelled dataset. As seen through this example, the LLM is capable of identifying three distinct intents in the request. Through its justification, it has correctly selected the appropriate portions of the request, contributing to the presence of the listed intents. One thing to know with this specific example is that the third intent is listed as a regular notification request. This intent could also have been interpreted as an intent report request. Based on the request, it is more likely that the user is referring to an intent report request rather than a regular notification request. Despite the similarity and the overlap between the two types of intents, additional clarification as to what constitutes one intent over the other is required. This clarification can be provided through additional prompt tuning or fine-tuning of the model with examples when it is converted into a conventional classification task. This model refinement will be further discussed and explored in future iterations of this work.

\begin{figure}[!htbp]
\centerline{\includegraphics[width=\columnwidth]{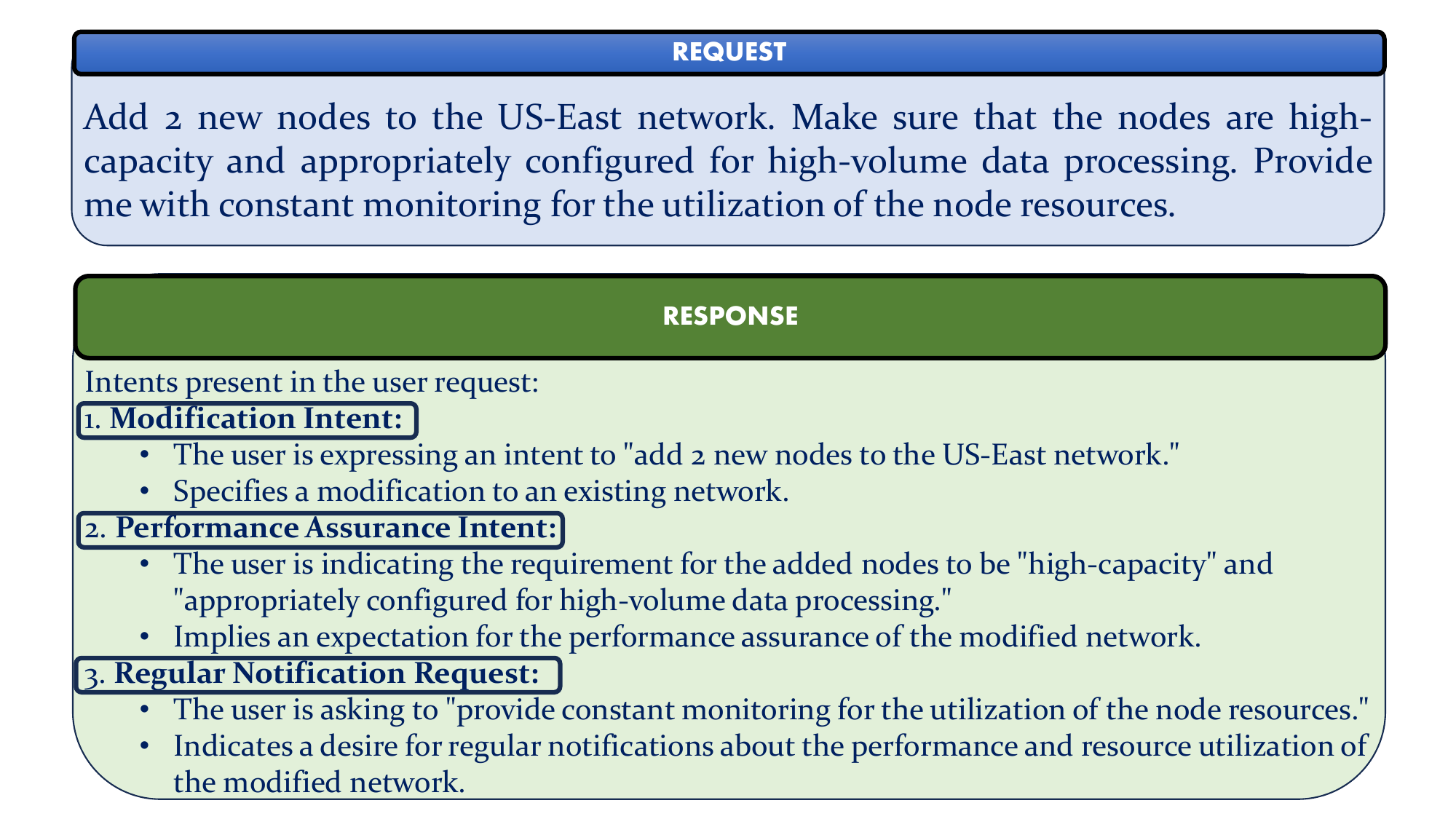}}
\caption{Multiple Intents Example 2}
\label{multi-2}
\end{figure}

Figure \ref{multi-2} illustrates another example of multiple intents present in a single request. In this modification request, performance assurance and regular notification are also present. The model's response demonstrates that it can accurately extract the various components of the request and associate them with their appropriate intents. Another thing to note for the future of this line of research is the ambiguity that may be present in specific requests due to the lack of information. For example, in terms of the regular notification request intent, no frequency is specified for the notifications. A genuinely autonomous end-to-end system should either request additional information from the user or assume a specific default value and notify the user of its choice.

\begin{figure}[!htbp]
\centerline{\includegraphics[width=\columnwidth]{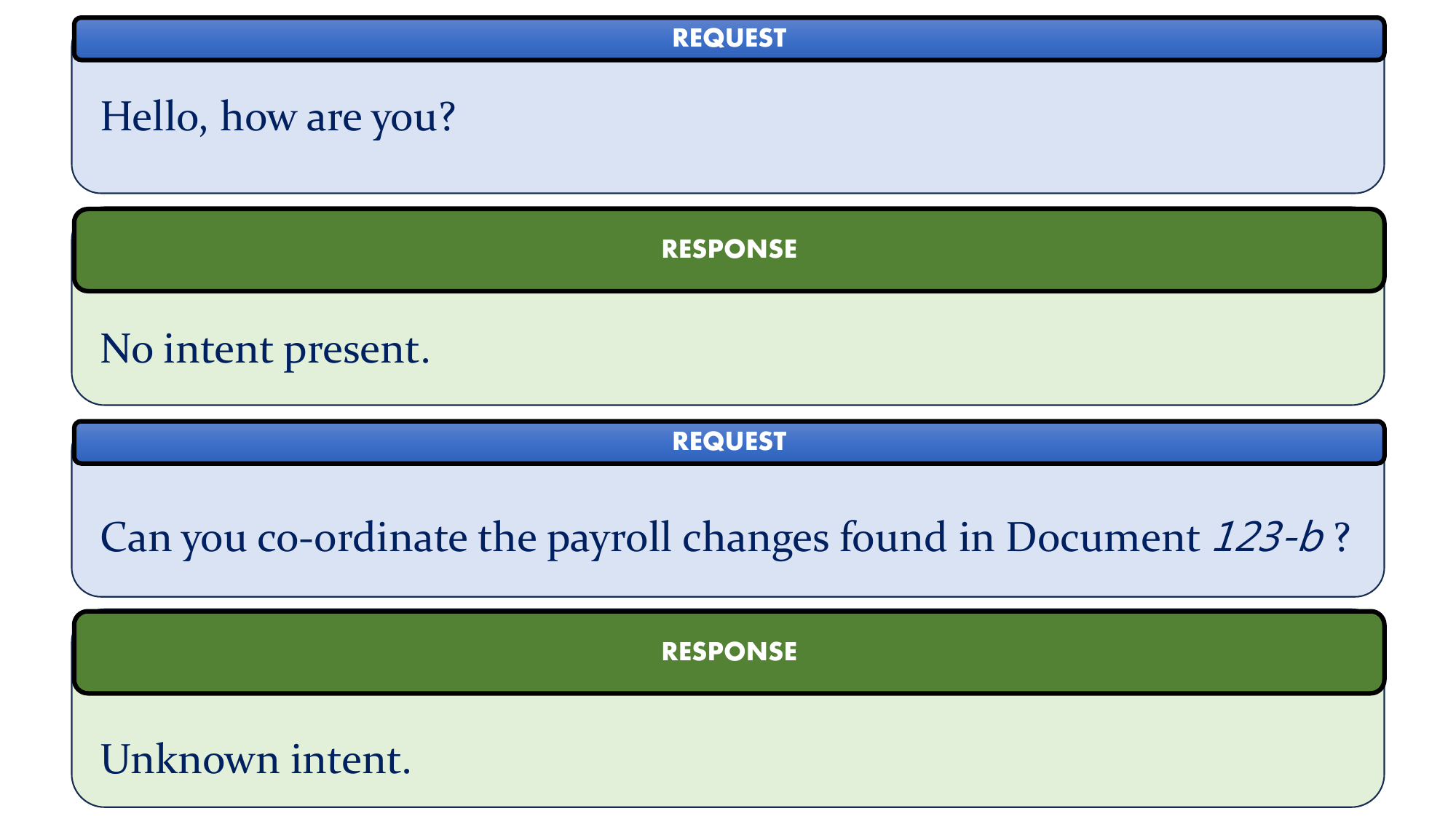}}
\caption{No Intent / Unknown Intent Examples}
\label{none}
\end{figure}

The final examples are presented in Figure \ref{none} and consist of requests that do not contain a specific intent or contain an intent unrelated to the mentioned intents of the 5G core network. In the first example, a conversational question is asked to the LLM. The model has appropriately determined that no intent is present, and based on the expected behaviour outlined in the prompt architecture, it has not answered the question but rather listed it as no intent present. In the second example, the request does contain a clear and specific intent; however, it is not related to the 5G core and cannot be classified into one of the six defined intent categories. Consequently, the LLM has appropriately listed this specific request as having an unknown intent. These examples show the versatility of the LLM and its ability to understand and interpret a variety of requests and classify them appropriately.

\section{Conclusion and Future Work}

In conclusion, intent-based networking is essential to traversing the road towards the creation and development of zero-touch networks. LLMs are a prime example of the latest research in the applications of intent extraction, interpretation, and conversion. The inputs and outputs of such models can be formatted for network management APIs in order to synthesize all functionalities pertaining to automated intelligence and analytics in next-generation networks. Future networks will rely on these advancements of artificial intelligence deployments to provide and further enhance optimal network performance and servicing for all network users and devices.\par

There are many avenues for future work stemming from the work presented in this paper. Firstly, regarding the LLM itself, future work will transition from a closed-source GPT-based model to an open-source state-of-the-art LLM, such as Llama 2 \cite{touvron2023llama}. Furthermore, future iterations of this work will move past resorting to prompting architecture exclusively and will include fine-tuning the model through custom datasets that are specific and comprehensive. This fine-tuning process will transform the task into a conventional classification task with a labelled dataset. A comprehensive and high-quality dataset representing the various intents required will be generated regarding this dataset. In order to generate this amount of data, LLMs will be leveraged along with data augmentation techniques to transform an initial set of seed prompts into a comprehensive and diverse dataset. Some text-based data augmentation techniques to be considered include back translation, paraphrasing, random erasing, and tone shift. Using these methods, diversity in text-based data will allow the model to learn the semantic meaning of the intent rather than simply memorizing a specific set or sequence of words and overfitting. Additionally, an analysis of the level of LLM hallucination will be conducted, and mitigation methods such as the popular Retrieval Augmented Generation (RAG) architecture will be explored. \par

The next avenue for future work includes actively integrating the developed LLM with a live 5G core testbed. Our previous work has extensively discussed and outlined the creation of an end-to-end 5G test bed with Core, RAN and Data Network capabilities \cite{chouman2022towards}. This testbed includes a custom Release 18 compliant implementation of the NWDAF, which will be the target deployment entity for this work. The NWDAF will have access to network-generated data \cite{manias2022nwdaf} and the ability to guide network management decisions \cite{10182447, 10255724}. By actively integrating this work with a live network, we can begin pushing the boundaries of innovation in order to attain a true autonomous end-to-end intelligent network. This integration will consist of developing the appropriate API calls within an architecture containing multiple LLMs communicating as part of a larger system to perform various services through decomposed micro functionalities such as intent extraction and entity extraction.\par

\bibliographystyle{IEEEtran}
\bibliography{sample}

\end{document}